\documentclass[12pt]{article}

\usepackage{xcolor}
\usepackage[sc]{mathpazo} % Use the Palatino font
\usepackage[T1]{fontenc} % Use 8-bit encoding that has 256 glyphs
\usepackage{microtype} % Slightly tweak font spacing for aesthetics

\usepackage[hmarginratio=1:1,top=32mm,columnsep=20pt]{geometry} % Document margins

\usepackage{amsmath,amssymb,amsfonts}
\usepackage{geometry}
\usepackage{amsthm}

\usepackage[utf8]{inputenc}
\usepackage{graphicx}
\usepackage{natbib}
\usepackage{hyperref}

\newtheorem{theorem}{Theorem}[section]
\usepackage{palatino}
\usepackage{float}
\title{\vspace{-15mm}\fontsize{22pt}{10pt}\selectfont\textbf{Spectral Curves with Complex Multiplication in Hermitian Matrix Models}} % Article title

\author{
\large
\text{Ali Nassar}%\thanks{A thank you or further information}
\\[2mm] % Your name
\normalsize Department of Physics, University of Science and Technology,\\ \normalsize Zewail City of Science and Technology,\\
\normalsize 12578 Giza, Egypt \\ % Your institution
\normalsize \href{mailto:anassar@zewailcity.edu.eg}{anassar@zewailcity.edu.eg} % Your email address
}
\date{\today}

\begin{document}

\maketitle

\begin{abstract}
We show that elliptic curves with complex multiplication (CM) naturally emerge in the spectral geometry of Hermitian one-matrix models in the two-cut phase. Focusing on a symmetric quartic potential, we derive the corresponding genus-one spectral curve and compute its modular $j$-invariant in  closed form as a function of the quartic coupling $g$. We identify specific values of $g$ for which the elliptic curve exhibits $CM$, i.e., its endomorphism ring is larger than $\mathbb{Z}$. This establishes  a direct  connection between number-theoretic structures and the spectral data of random matrix ensembles.
\end{abstract}

\section{Introduction}
\label{sec:intro}

Random matrix models provide a unifying framework for a wide range of problems in physics and mathematics, from quantum gravity and string theory to number theory, statistical mechanics, and quantum chaos \cite{Mehta2004,GinspargMoore1993,Bourgade2013,Haake2001,DiFrancesco1995,Klebanov:1991qa}. Random matrix ensembles of large dimension were first introduced by Wigner in 1951~\cite{Wigner51} 
to study the spectral properties of complex systems exhibiting chaotic behavior. In this framework, the Hamiltonian of a chaotic system is modeled as a large matrix with random entries. One of the most remarkable aspects of matrix models is the emergence of
a spectral curve, capturing the large-$N$ behavior of eigenvalues and encoding the analytic structure of the theory in geometric terms
\cite{Brezin:1977sv,EynardOrantin2007,Eynard1992,Eynard2004} (see~\cite{Eynard:2015aea} for a detailed account). In particular, multi-cut solutions of matrix models give rise to higher-genus spectral curves, with the two-cut phase corresponding to a genus-one elliptic curve.

In this work, we explore how spectral curves can exhibit complex multiplication (CM), a deep arithmetic property of elliptic curves \cite{ParshinShafarevich1992,Cox1989,ShimuraTaniyama1961,Lang1991}
(For a comprehensive introduction, as well as applications of number‑theoretic ideas in string theory, see~\cite{Moore:1998pn}). For the Hermitian matrix model with quartic potential  
\begin{equation}
V(x) = -\tfrac{1}{2}x^2 + \tfrac{g}{4}x^4,
\end{equation}  
we focus on the range of $g$ that produces a two-cut phase, corresponding to a genus-one spectral curve. We explicitly determine this curve and compute its $j(g)$-invariant as a function of $g$. In particular, we find ten admissible values of $g$ at which the spectral curve acquires complex multiplication, meaning its endomorphism ring is larger than $\mathbb{Z}$. At these special values of $g$, the automorphism group of the curve is enhanced. For instance, for $j=1728$ the usual group of symmetries of the elliptic curve which is of order $2$  is promoted to a symmetry group of order $4$, with similar enhancements occurring at the other CM points.

We find that, for every complex multiplication value of the $j$-invariant $j_{\mathrm{CM}}$, the equation
$
j(g)=j_{\mathrm{CM}}
$
possesses two distinct solutions for the coupling $g$, defining two branches $g_{-}(j)$ and $g_{+}(j)$. The larger branch $g_{+}(j)$ converges rapidly to the critical coupling $g_c=1/4$, which marks the double-scaling limit where the associated spectral curve degenerates.

Elliptic curves with complex multiplication have been studied by various authors in different contexts. 
For example, in~\cite{Gukov:2002nw,Hosono:2002yb} the authors formulated a criterion for the rationality of 
$c=2$ conformal field theories on an elliptic curve $E$, with the 
essential requirement that $E$ possess complex multiplication.

From a physical perspective, the emergence of Complex Multiplication (CM) in the spectral geometry of matrix models is of significant importance due to the deep connections between matrix models, topological string theory, and supersymmetric gauge theories (see \cite{Marino:2004eq} and references therein). In these contexts, the spectral curve of the matrix model is identified with the mirror curve of a local Calabi-Yau manifold, the Seiberg-Witten curve of a gauge theory, or the macroscopic loop equation of 2D quantum gravity. The points in the moduli space where this spectral curve acquires CM correspond to special physical vacua where the theory exhibits enhanced symmetries.

The organization of this paper is as follows. In Section \ref{sec:2}, we review the basics of Hermitian matrix models and their spectral curves in the large‑$N$ limit. Section \ref{sec:3} develops the construction of the spectral curve and introduces its genus‑one form in the two‑cut phase. In Section \ref{sec:3.1}, we provide a concise exposition of elliptic curves, focusing on the algebraic formulation needed for our analysis, while Section \ref{sec:3.2} discusses the notion of complex multiplication. Section \ref{sec:4} contains the main results: we compute the $j$-invariant of the spectral curve in the quartic matrix model and identify the special values of the coupling constant for which complex multiplication occurs. Finally, Section \ref{sec:5} summarizes our findings and outlines possible directions for future research.

\section{Hermitian Matrix Models}\label{sec:2}

Random matrix theory studies ensembles of matrices whose entries are drawn from a specified probability distribution. Here we will consider the Gaussian Unitary Ensemble (GUE) of Hermitian $N\times N$ matrices $M$.
 The matrix elements $M_{ij}$ are chosen according to the following probability distribution:
\begin{equation}
P(M_{ii}) = \frac{1}{\sqrt{\pi}} e^{-\frac{1}{2}M_{ii}^2},\qquad P(M_{ij}) = \frac{1}{\pi} e^{-|M_{ij}|^2},\quad i<j.
\end{equation}
The probability of choosing a matrix $M$ is therefore given by
\begin{equation}
P(M)=\prod_{i=1}^N P(M_{ii}) \prod_{1\leq i< j\leq N }^N P(M_{ij}) =A_N \prod_{i, j=1 }^N  e^{-\frac{1}{2}|M_{ij}|^2}=A_N e^{-\frac{1}{2}\text{Tr}(M^2)},
\end{equation}
where $A_N$ is a constant.

  The partition function is
\begin{equation}
 Z= \int_{\mathcal{H}_N}  d M e^{-\frac{1}{2}\text{Tr}(M^2)},   
\end{equation}
  where $\mathcal{H}_N$ is the space of $N \times N$ Hermitian matrices. The Lebesgue measure \( dM \) on the space \( \mathcal{H}_N \) is given by:
\begin{equation}
dM = \prod_{i=1}^N dM_{ii} \prod_{1 \le i < j \le N} d\,\Re M_{ij} \, d\,\Im M_{ij},
\end{equation}
where $\Re M_{ij} \, d\,\Im M_{ij}$ denote the real and imaginary parts of $M_{ij}$, respectively. The measure $dM$ is invariant under unitary conjugation. 

We can generalize the preceding construction by replacing the quadratic term $M^2/2$ with a general potential $V(M)$, where $V$ is a polynomial (or, more generally, analytic function). The Gaussian matrix model, characterized by the quadratic potential $V(M)=M^2/2$, is the simplest example in random matrix theory. It serves as a starting point for understanding the universal features of eigenvalue statistics, such as Wigner's semicircle law \cite{Wigner:1967qdh}. Moreover, it admits a complete analytic solution and establishes connections to various physical systems, such as quantum chaos, nuclear spectra, and statistical mechanics. However, it corresponds to a single-cut solution in the large-$N$ limit and fails to capture the richer phase structure or critical behavior present in more complex systems.

To explore phenomena such as multi-cut solutions, phase transitions, and connections to two-dimensional quantum gravity or string theory, it is necessary to go beyond the Gaussian  potential and consider higher degree polynomials, for example,
\begin{equation}
 V(M)=\sum_k \frac{g_k}{k} M^k,   
\end{equation}
where $g_k$ are  coupling constants. These deformations introduce new critical points and allow for the realization of multiple eigenvalue supports (multi-cut phases) \cite{Brezin:1977sv,Eynard:2003zu}. Such generalizations are crucial for accessing non-trivial universality classes, studying double-scaling limits, and constructing a bridge between matrix models and continuum theories of random surfaces \cite{Brezin:1990rb,Douglas:1989ve,Gross:1989ni,Gross:1989aw}. 

We consider the partition function
\begin{equation}\label{eq:Zpart}
 Z_N= \int_{\mathcal{H}_N}  d M e^{-N\text{Tr}V(M)},   
\end{equation}
where we have introduced an explicit factor of $N$ in the exponent of the matrix integral. This normalization is motivated by the large-$N$ limit, where $N$ plays a role analogous to $1/\hbar$ in quantum mechanics or inverse temperature in statistical mechanics. 
This normalization also guarantees that the eigenvalues of $M$ remain of order one and do not scale with $N$ in the large-$N$ limit.

To analyze the partition function (\ref{eq:Zpart}), we use the fact that $M$ can be diagonalized by a unitary transformation $M = U \Lambda U^\dagger$, with $\Lambda = \mathrm{diag}(\lambda_1, \dots, \lambda_N) )$. The measure becomes 
\begin{equation}
dM =  \prod_{i=1}^N d\lambda_i \prod_{1 \le m < n \le N} (\lambda_m - \lambda_n)^2\, d\mu(U),
\end{equation}
where $d\mu(U)$ is the Haar measure on the unitary group  $U(N)$. The eigenvalue representation of the partition function is
\begin{equation}
Z_N =C_N \int_{\mathbb{R}^N} \prod_{i=1}^N d\lambda_i\, \prod_{1 \le i < j \le N} (\lambda_i - \lambda_j)^2\, e^{-N \sum_{i=1}^N V(\lambda_i)},
\end{equation}
where $C_N$ is a normalization constant that will play no role in the subsequent analysis and will therefore be omitted.

We can express $Z_N$ in terms of an effective action:
\begin{equation}
Z_N = \int_{\mathbb{R}^N} \prod_{i=1}^N d\lambda_i \, e^{- N^2 S_{\text{eff}}(\lambda_1, \cdots, \lambda_N)},
\end{equation}
where
\begin{equation}
 S_{\text{eff}}(\lambda_1, \dots, \lambda_N) = \frac{1}{N}\sum_{i=1}^N V(\lambda_i) - \frac{1}{N^2} \sum_{1 \le i < j \le N} \log|\lambda_i - \lambda_j|.
\end{equation}

In the large-$N$ limit, the partition function  is dominated by configurations that minimize the effective action $S_{\text{eff}}$. This motivates the use of the saddle-point  approximation to evaluate the matrix integral. 
The saddle-point equations are
\begin{equation}
\frac{\partial S_{\text{eff}}}{\partial \lambda_i} =0,
\end{equation}
which leads to
\begin{equation}\label{eq:Saddle}
V'(\lambda_i) = \frac{2}{N} \sum_{j \ne i} \frac{1}{\lambda_i - \lambda_j}.
\end{equation}

To solve the saddle point equations, we define the microscopic eigenvalue density.
\begin{equation}\label{eq:rho}
\rho(\lambda) = \frac{1}{N} \sum_{i=1}^N \delta(\lambda - \lambda_i),\qquad \int \rho(\lambda) =1.
\end{equation}
We also define the resolvent:
\begin{equation}
W_N(x) =\frac{1}{N}  \mathrm{Tr} \frac{1}{x - M} =  \frac{1}{N} \sum_{i=1}^N  \frac{1}{x - \lambda_i},\qquad x\in \mathbb{C}.
\end{equation}
The above equation can be written as: 
\begin{equation}
W(x) =\int \frac{\rho(\lambda)}{x-\lambda} d\lambda.
\end{equation}
This shows that the resolvent is  the Stieltjes transform of $\rho(\lambda)$. For finite $N$, the resolvent defines a meromorphic function on $\mathbb{C}$ with simple poles at $x=\lambda_i \in \text{Supp}\, \rho $, all on the real axis since $M$ is Hermitian.

From the above definitions one can show that $W(x)$ satisfies 
\begin{equation}
W(x)^2 + \frac{1}{N} W'(x)
= \frac{2}{N^2} \sum_{i=1}^N \frac{1}{x - \lambda_i} \sum_{j (\neq i)} \frac{1}{\lambda_i - \lambda_j}
= \frac{1}{N} \sum_{i=1}^N \frac{V'(\lambda_i)}{x - \lambda_i}. 
\end{equation}
Introducing
\begin{equation}\label{barP}
P(x) = \frac{1}{N} \sum_{i=1}^N \frac{V'(x) - V'(\lambda_i)}{x - \lambda_i},
\end{equation}
we obtain the Riccati equation for the resolvent \cite{Brezin:1977sv,Eynard:2015aea}
\begin{equation}\label{eq:Riccati}
W(x)^2 + \frac{1}{N} W'(x) = V'(x) W(x) - P(x).
\end{equation}

Now let us go back and  study the saddle point equations (\ref{eq:Saddle}).
In the limit $N\rightarrow \infty$, we assume that the eigenvalues will condense into a compact support $\mathcal{C}$ on the real axis. In general, the support $\mathcal{C}$ consists of 
 a finite union of disjoint intervals, commonly referred to as cuts
\begin{equation}
\mathcal{C}=\bigcup_{i=1}^s [a_{2i-1},a_{2i}].
\end{equation}
The density $\rho$ can be approximated by a continuous non-negative function\footnote{This differs from the microscopic density in (\ref{eq:rho}), but we will nevertheless denote it by $\rho$.} $\rho$ on $\mathcal{C}$.

In the large-$N$ limit, the simple poles of the resolvent coalesce into branch cuts. 
The resolvent  $W(x)$ becomes an analytic function on the complex plane, 
except along the support $ \mathcal{C}$ of the eigenvalue density, 
where it develops branch cut  discontinuities.
Using Sokhotski–Plemelj formula
\begin{equation}
\lim_{\epsilon \to 0^+} W(\lambda \pm i\epsilon) 
= \mathrm{P.V.} \int_{\mathcal{C}} \frac{\rho(\lambda')}{\lambda - \lambda'} \, d\lambda' 
\mp i\pi  \rho(\lambda).
\end{equation}
This gives the relations:
\begin{equation}\label{eq:Sok-Plem}
\rho(\lambda) = \frac{1}{2\pi i} \left[ W(\lambda - i0) - W(\lambda + i0) \right],\qquad 
 \frac{1}{2} V'(\lambda)=W(\lambda + i0) + W(\lambda - i0).
\end{equation}
Thus, to determine the eigenvalue density $\rho$ in the large-$N$ limit, we must first compute the resolvent $W(x)$ and evaluate its discontinuity across the real axis. For this, we go back to Riccati equation  (\ref{eq:Riccati}) which in the large-$N$ limit becomes an algebraic equation for the resolvent\footnote{We will continue to use the symbols $W$ and $P$, even though what is meant are their large‑$N$ counterparts.}
\begin{equation}\label{eq:Riccati2}
W(x)^2  = V'(x) W(x) - P(x).
\end{equation}
%where 
%\begin{equation}\label{eq:resolvent}
%\overline{W}(x)  = \lim_{N\mapsto \infty } W(x),%\qquad 
%\overline{P}(x)  = \lim_{N\mapsto \infty } P(x)
%\end{equation}
The solution is
\begin{equation}\label{eq:resolvent}
W(x)  = \frac{1}{2}\Big(V'(x) -\sqrt{V'(x)^2-4 P (x)} \Big).
\end{equation}
Let us assume that $V(x)$ is a polynomial of degree $D+1$ and recall from (\ref{barP}) that $P$ is a polynomial of degree $D-1$. Then $(V')^2-4P$ is a polynomial of degree $2D$. The polynomial $(V')^2-4P$ can be factorised into even and odd zeros
\begin{equation}\label{eq:Msigma}
(V')^2-4P= M^2 \sigma,
\end{equation}
where all the zeros of $\sigma$ are simple. 
Since the degree of $(V')^2-4P$ is even then $\sigma$ is even degree polynomial
\begin{equation}
\sigma(x) = \prod_{i=1}^{2s} (x-a_i).
\end{equation}

By substituting (\ref{eq:Msigma}) into (\ref{eq:Sok-Plem}) and computing the discontinuity across the real $\lambda$-axis, we obtain 
\begin{equation}\label{rhofromW}
\rho(\lambda)= M(\lambda)\sqrt{ \sigma(\lambda)},\quad \lambda\in \mathcal{C}.
\end{equation}
This shows that  $\rho(\lambda)$ is supported on $s$ intervals and we call $s$ the number of cuts.

We can use (\ref{rhofromW}), along with the requirement that $ \rho(\lambda)$  is non-negative on $ \mathcal{C} $, to put an upper bound on the number of cuts $s$. Since $ \sqrt{\sigma(\lambda)}$  changes sign when moving from one interval to the next, $ M(\lambda)$ must also change sign across these intervals. This implies that $ M(\lambda)$  has at least one zero in each of the  $s-1 $ regions separating the $ s $ intervals. Therefore,
\begin{equation}\label{eq:DegM}
\text{Deg}(M) \geq s-1.
\end{equation}
From (\ref{eq:Msigma}),
\begin{equation}\label{eq:DegM2}
\text{Deg}(M)=D-s.
\end{equation}
Hence,
\begin{equation}\label{eq:numCuts}
s\leq \frac{D+1}{2}.
\end{equation}

\section{The Spectral Curve}\label{sec:3}
In this section, we derive the equation of the spectral curve associated with a matrix model defined by a potential $V$. As we have seen, in the large-$N$ limit, the resolvent satisfies a single algebraic equation\begin{equation}
W(x)^2  - V'(x) W(x) + P(x)=0.
\end{equation}
Defining
\begin{equation}
y=V'(x)-2W(x).
\end{equation}
We obtain
\begin{equation}\label{eq:rsurface}
y^2 = V'(x)^2 - 4 P(x) = \mathcal{F}(x).
\end{equation}
This equation defines an algebraic curve $\Sigma \subset \mathbb{C} \times \mathbb{C}$ given by
\begin{equation}
\Sigma := \left\{ (x, y) \in \mathbb{C} \times \mathbb{C} \ \big| \ y^2 =\mathcal{F}(x)  \right\}.
\end{equation}
The genus \footnote{What we are really talking about here is the associated compact Riemann surface which is obtained from (\ref{eq:rsurface}) by resolving the singularities and adding a point or two to make it compact.} of $\Sigma$ is determined by the degree and root structure of the polynomial $\mathcal{F}(x)$. If $\mathcal{F}(x)$ is of degree $\mathcal{N}$ and has distinct roots, then the genus $h$ of the  curve satisfies
\begin{equation}
\mathcal{N} = 2h + 1 \quad \text{or} \quad \mathcal{N} = 2h + 2.
\end{equation}
If $V(x)$ is a polynomial of degree $D+1$, then $\mathcal{F}(x)$  has degree $2D$, leading to a genus $h = D - 1$. For example, a quartic potential $V(x)$ generically yields a genus-two hyperelliptic curve. However, if $\mathcal{F}(x)$ has repeated roots, then the curve $\Sigma$ becomes singular and its genus is reduced accordingly due to degeneracy.

To determine the algebraic curve associated with a given matrix model, we begin with the expression for the resolvent (\ref{eq:resolvent}) which can now be written as
\begin{equation}
y = M(x)\sqrt{\sigma(x)}.
\end{equation}
 Since $M(x)$ is a polynomial, we can perform a birational transformation $y \to y/M(x)$, which yields an isomorphic curve\footnote{We emphasize that the transformation {\color{red}{$y\to y/M(x)$}}  is used to isolate the underlying complex structure of the curve. Because this is a birational transformation, it leaves the complex structure and the modular $j$-invariant  unchanged.  Consequently, for the arithmetic classification considered here, it is sufficient to work with the simplified curve (\ref{eq:NoMquartic}).}
\begin{equation}\label{eq:NoMquartic}
y^2 = \prod_{i=1}^{2s} (x - a_i).
\end{equation}
The right-hand side is a degree $2s$ polynomial, so the resulting curve has genus 
\begin{equation}
h = s - 1.    
\end{equation}
 %{\color{red}{Geometrically, this describes a double cover of $%\mathbb{C}$, branched over the $2s$ points $a_i$.}}
 Since $y = V' - 2W$ and $V'$ is a polynomial, the discontinuity of $y$ determines the discontinuity of $W$, which in turn determines the spectral density $\rho$. Therefore, the algebraic curve $y^2=\mathcal{F}(x)$ encodes the large-$N$ spectrum, and in this sense, it is referred to as the spectral curve. However, the spectral curve encodes more than just the spectrum; it also serves as the foundation for topological recursion  \cite{Chekhov:2006vd,EynardOrantin2007,Eynard2004}. This transforms the problem of solving the loop equations  into a universal algorithm based solely on the local geometry of the spectral curve. In this way, the data of the matrix model like higher-genus contributions to the free energy and correlation functions is  fully encoded in the geometry of the curve. Our main interest will be the two-cut matrix models, $s=2$, which gives a  genus-one curve, that is, an elliptic curve defined by the equation 
 \begin{equation}\label{eq:ellipCurve}
y^2 =  (x - a_1)(x - a_2)(x - a_3)(x - a_4).
\end{equation}
The spectral edges $a_i$ for a given potential $V(x)$ are determined from the large $x$ asymptotics of the resolvent together with the normalization condition, as will be detailed in the next section.

\subsection{Elliptic Curves}\label{sec:3.1}

In this section, we provide a brief exposition of elliptic curves, focusing only on the concepts that will be needed in the rest of the paper. Elliptic curves can be  realized as  plane algebraic curves defined by a cubic equation in two variables. This perspective not only makes the connection with spectral curves more transparent, but also provides the minimal background necessary for the results we will present.   We can take (\ref{eq:ellipCurve}) and convert it into Weierstrass normal form (see Appendix (\ref{appendix}))
\begin{equation}\label{ellipcurve}
y^2=4x^3-g_2x-g_3, \quad x,y\in \mathbb{C},
\end{equation}
where $g_2$ and $g_3$ are complex numbers.  

The curve (\ref{ellipcurve}) is non-singular if there are no points $(x,y)$  satisfying $f(x,y)=0$ and $\nabla f(x,y)=0$ simultaneously, where $f$ is the polynomial defining the curve. For (\ref{ellipcurve}),  this translates to
\begin{equation}
\Delta=g_2^3-27g_3^2\neq 0.
\end{equation}
The polynomial $\Delta$ is known as the \textit{discriminant} of the elliptic curve. Over $\mathbb{C}$, every such curve can be mapped to a complex torus.

A complex torus $T^2$  can be constructed in the following way: start with a lattice $\Lambda$ in the complex plane
\begin{equation}
\Lambda = \mathbb{Z}  \omega_1 \oplus \mathbb{Z} \omega_2,
\end{equation}
where $\{\omega_1,\omega_2\}$  are the basis vectors of the lattice. Then
a complex torus is the quotient of the complex plane by the lattice $\Lambda$
\begin{equation}
\mathbb{C}/\Lambda =\{z \sim z+\Lambda : z\in \mathbb{C} \}.
\end{equation}
By rescaling we can bring the lattice to  the form
\begin{equation}
\Lambda = m  1 + n \tau,\quad \tau =\frac{\omega_1}{\omega_2}  \in \mathbb{H}^+,
\end{equation}
where $\mathbb{H}^+$ is the upper-half  plane.  The $\text{SL}(2,\mathbb{Z})$ action on $\omega_1$ and $\omega_2$ induces the following $\text{SL}(2,\mathbb{Z})$ action on $\tau$ 
\begin{equation}
\tau \mapsto \frac{a\tau+b}{c\tau+d} \qquad \text{with}\quad 
\begin{pmatrix}
a&b\\
c&d
\end{pmatrix}
\in SL(2,\mathbb{Z}), \quad ab-cd=1,
\end{equation}
which gives an equivalent torus. Since the above action on $\tau$ remains the same if we  change the signs of $a,b,c,d$, then only $\text{PSL}(2,\mathbb{Z})=\text{SL}(2,\mathbb{Z})/\mathbb{Z}_2$ acts faithfully. The group   
$\text{PSL}(2,\mathbb{Z})$ is the \textit{modular group} of the torus.  The modular group is generated by the  two operations 
\begin{equation}
\begin{split}
 T: \tau\longrightarrow \tau+1 \quad  \text{or }   
 T=\begin{pmatrix}
 1 &1\\
 0&1
 \end{pmatrix}\\
  S: \tau\longrightarrow -\frac{1}{\tau}
   \quad  \text{or }   
S=\begin{pmatrix}
0 &-1\\
 1&0
 \end{pmatrix}.
\end{split}
\end{equation}
The set of inequivalent tori is parametrized by $\mathbb{H}^+$ modulo the $T$ and $S$ transformations. The fundamental domain of $\tau$ is  
\begin{equation}
\mathcal{F}_0=\Big\{ -\frac{1}{2}<\mathcal{R}(\tau)\leq \frac{1}{2},\ \Im(\tau)>0,\ |\tau|\geq 1   \Big\}.
\end{equation}
The values of $\tau\in \mathcal{F}_0$ parametrize inequivalent tori, i.e., tori which cannot be transformed into one another by $\text{PSL}(2,\mathbb{Z})$.

To establish an isomorphism between elliptic curves and complex tori, we need to define the Weierstrass $\wp$-function.  First, an elliptic function is a function on $\mathbb{C}$  which is doubly periodic for $\Lambda$
\begin{equation}
\begin{split}
F(z+\omega_1)&=F(z)\\
F(z+\omega_2)&=F(z).
\end{split}
\end{equation}
The Weierstrass $\wp$-function is an elliptic function defined as
\begin{equation}
\wp (z) = \frac{1}{z^2}+\sum_{\omega\in \Lambda}{}^{'}\Big[\frac{1}{(z-\omega)^2} - \frac{1}{\omega^2} \Big],
\end{equation} 
where $\sum'$ means we only sum over non-zero lattice vectors. Using the notation
\begin{equation}
s_m(\Lambda)=\sum_{\omega\in \Lambda}{}^{'} \frac{1}{\omega^m}
\end{equation}
we get the expansion
\begin{equation}
\wp (z) = \frac{1}{z^2}+\sum_{n=1}^\infty (2n+1) s_{2n+2}(\Lambda)z^{2n}.
\end{equation} 
The first few terms are
\begin{equation}
\wp (z) = \frac{1}{z^2}+3s_4(\Lambda)z^{2}+5s_6(\Lambda)z^{4}.
\end{equation} 

If we define $g_2=g_2(\Lambda)=60s_4$ and $g_3=g_3(\Lambda)=140s_6$. Then one can show that
\begin{equation}
\wp'^2=4\wp^3-g_2 \wp-g_3.
\end{equation}
Thus the points $(x,y)=(\wp(z),\wp'(z))$ lie on the elliptic curve
\begin{equation}
y^2=4x^3-g_2 x-g_3.
\end{equation}
It can be shown that the discriminant of this elliptic curve $\Delta(\Lambda)$ is non-zero,  ensuring that the equation indeed defines a nonsingular elliptic curve. This leads to the following uniformization theorem

\begin{theorem}[Uniformization theorem]
Let $E$ be an elliptic curve given by the Weierstrass equation
\begin{equation}
y^2 = 4x^3 - g_2 x - g_3,
\end{equation}
where $g_2, g_3 \in \mathbb{C}$ and $g_2^3 - 27 g_3^2 \neq 0$. 
Then there exists a unique lattice $\Lambda$ such that
\begin{equation}
g_2 = g_2(\Lambda), \qquad g_3 = g_3(\Lambda).
\end{equation}
\end{theorem}
The isomorphism between complex tori and elliptic curves can be written as
\begin{equation}
\begin{gathered}
\psi \  : \ \mathbb{C}/(\mathbb{Z}+\tau \mathbb{Z})\longrightarrow E\\
z \mapsto \psi(z) = [\wp (z), \wp' (z)].
\end{gathered}
\end{equation}

\subsection{Complex Multiplication}\label{sec:3.2}

Consider an elliptic curve (or a torus)
%complex number $\tau=\tau_1+i\tau_2$ with $\tau_2>0$ we can construct an elliptic curve
 $E_\tau=\mathbb{C}/\Lambda $, where $\Lambda$ is a lattice  $\Lambda=(\mathbb{Z}+\tau\mathbb{Z})$.
 %the  of $E_\tau$. The elliptic curve $E_\tau$ describes a torus which is given by identifying points in the complex plane related %by the lattice translations generated by $1$ and $\tau$.
The endomorphisms of $E_\tau$ are given by holomorphic maps $ F: E_\tau\rightarrow E_\tau$.
  The only such holomorphic maps fixing $0$ are induced by maps $G: \mathbb{C}\rightarrow \mathbb{C}$ of the form $ G(z)=\lambda z$ where $\lambda\in\mathbb{C}$ is a constant such that $\lambda \Lambda \subseteq \Lambda$. Any elliptic curve will admit  a set of trivial endomorphisms corresponding to multiplication by  integers $\lambda \in \mathbb{Z}$, since multiplication by integers moves one from one lattice point to another. However, some special elliptic curves could have non-trivial endomorphisms for particular values of $\tau$.  We want to find the conditions on $\tau$ which give non-trivial endomorphisms.

To characterize the possible endomorphisms $\mathrm{End}(E_\tau)$ of $E_\tau$, 
we consider how multiplication by a complex number $\lambda$ acts on the 
lattice generators. For $\lambda \in \mathrm{End}(E_\tau)$, there must exist 
integers $j,k,m,n \in \mathbb{Z}$ such that the action of $\lambda$ on the 
lattice $\Lambda$ is expressed on its generators as
\begin{equation}\label{endo}
\lambda \cdot 1= j \cdot 1+k \tau,\qquad 
\lambda \tau= m \cdot 1+n \tau, \quad j,k,m,n \in\mathbb{Z}.
\end{equation}
Writing (\ref{endo}) as
\begin{equation}
\begin{pmatrix}
\lambda - j &  -k \\
- m & \lambda- n\\
\end{pmatrix}
\begin{pmatrix}
 1 \\
 \tau
\end{pmatrix}
=
\begin{pmatrix}
0 \\
 0
\end{pmatrix}.
\end{equation}
The existence of non-trivial solutions of the above equation implies that
\begin{equation}
\lambda^2-(j+n) \lambda  +(jn-km)=0,
\end{equation}

If $\lambda \in \mathbb{R}$, and since $1$ and $\tau$ are linearly independent over $\mathbb{R}$, the first relation in~\eqref{endo},
\begin{equation}\label{lin-ind}
(\lambda - j)\cdot 1 - k\tau = 0,
\end{equation}
forces $\lambda = j$. Hence $\lambda \in \mathbb{Z}$, which corresponds to the trivial endomorphisms of any elliptic curve.
If, on the other hand, $\lambda \in \mathbb{C}$, then  
\begin{equation}\label{betaaaa}
\lambda =\frac{(j+n)\pm \sqrt{(j-n)^2+4km} }{2}.
\end{equation}
Using this value of $\lambda$ in (\ref{lin-ind})  we  get
\begin{equation}
\tau =\frac{(n-j)+ \sqrt{(j-n)^2+4km} }{2k}.
\end{equation}
Define $b=j-n$, $c=-m$, and $a=k$ and dividing by the greatest common factor of $b=j-n$, $c=m$, and $a=k$ if necessary then we can write  
\begin{equation}
\tau =\frac{-b+ \sqrt{b^2-4ac} }{2a}=\frac{-b+ \sqrt{D} }{2a},
\end{equation}
where $D=b^2-4ac<0$. A complex number of the form 
\begin{equation}
\tau = \alpha + \beta \sqrt{D}, \qquad \alpha, \beta \in \mathbb{Q}, \; D<0,
\end{equation}
is said to lie in the imaginary quadratic number field  $\tau\in \mathbb{Q}(D)$.

We conclude that for a generic value of $\tau$, the endomorphism ring of the elliptic curve, $\operatorname{End}(E_\tau)$,
consists only of multiplications by integers.  
However, if $\tau$ lies in an imaginary quadratic field, $\tau \in \mathbb{Q}(D)$, then $E_\tau$ admits a larger endomorphism ring, generated by multiplications with elements of the form~\eqref{betaaaa}.  
Elliptic curves (or tori) corresponding to these special values of $\tau$ are said to have \emph{complex multiplication}, or equivalently, to be of \emph{CM type}.

We end this section with what is called \textit{the first main theorem of complex multiplication} {{\color{red}{\cite{silverman2009arithmetic}}}. This theorem says that for $\tau\in \mathbb{Q}(D)$, then $j(\tau)$ is an algebraic integer, i.e.,
\begin{equation}\label{eq:algebint}
j^h+a_1j^{h-1}+\cdots+a_h=0, \quad a_i\in \mathbb{Z},
\end{equation}
where $h=h(D)$ is the class number of the field $\mathbb{Q}(D)$ \cite{ParshinShafarevich1992,ConwaySloane1993}. We see that although $j(\tau)$ is a non-trivial function of $\tau$, for elliptic curves with complex multiplication $j(\tau)$ has a very simple property. 
Note that the fact that $j$ is an algebraic integer provides a necessary condition for an elliptic curve to have complex multiplication, but it is not by itself sufficient.

\section{Spectral curves with CM}\label{sec:4}
In this section we compute the $j$-invariant of the elliptic curve arising as 
the spectral curve of the two-cut quartic matrix model. Since genus one curves 
correspond to two cuts, by $h = s - 1$ and using (\ref{eq:numCuts}), the 
potential must have degree at least four. We therefore consider the quartic 
potential as the simplest case admitting a two-cut solution
\begin{equation}
 V(x)=-\frac{1}{2}x^2 +\frac{g}{4}x^4, \qquad g>0. 
\end{equation}
The potential has two minima at $x=\pm \sqrt{2/g}$ and a maximum at the origin. According to (\ref{eq:numCuts}), the system can exist in a single-cut phase or a two-cut phase. By tuning $g$, we can reach the two-cut regime. This occurs when the minima are sufficiently deep that the eigenvalues cluster around each minimum independently, without tunneling between them. To find the spectral edges, we assume symmetric cuts  $\mathcal{C}= [-a,-b]\cup[b,a]$. The resolvent becomes
\begin{equation}
 W(x)=\frac{1}{2}(V'(x)-M(x)\sqrt{(x^2-a^2)(x^2-b^2)}).
\end{equation}
From (\ref{eq:DegM2}), $\text{Deg}(M)=D-s=1$. Therefore
\begin{equation}\label{eq:Res2}
 W(x)=\frac{1}{2}(-x+gx^3-(\alpha+\beta x)\sqrt{(x^2-a^2)(x^2-b^2)}).
\end{equation}
To find $\alpha,\beta$, we use the asymptotic behavior of $W(x)$ at infinity:
\begin{equation}
\lim_{x\rightarrow \infty} W(x)\sim\frac{1}{x}.
\end{equation}
Expanding (\ref{eq:Res2}) around $x\rightarrow \infty$
\begin{equation*}
W(x)
= \frac{1}{2}(g - \beta)x^{3} - \frac{\alpha\, x^{2}}{2} + \frac{1}{4}\big(-2 + a^{2}\beta + b^{2}\beta\big)x
+ \frac{1}{4}(a^{2} + b^{2})\alpha + \frac{(a^{2} - b^{2})^{2}\beta}{16 x}+\mathcal{O}\bigg[\frac{1}{x^2}\bigg].
\end{equation*}
Hence, we get the following conditions
\begin{equation}
\alpha=0, \qquad \beta=g,\qquad a^2+b^2=\frac{2}{g},\qquad (a^2-b^2)^2=\frac{16}{g}.
\end{equation}
The last two equations give the spectral edges
\begin{equation}\label{SpecEdges}
a^2=\frac{1}{g}+\frac{2}{\sqrt{g}}, \qquad b^2=\frac{1}{g}-\frac{2}{\sqrt{g}},  \quad a>b
\end{equation}
The two-cut phase occurs for $0<g<1/4$. As $g\rightarrow 1/4$, the gap between the two cuts closes and we get a single cut phase.  The critical value $g_c=1/4$

Using the spectral edges we found in (\ref{SpecEdges}), the equation of the curve becomes
\begin{equation}\label{eq:quartic}
y^2=(x^2-a^2)(x^2-b^2).
\end{equation}
This equation can be converted to the Legendre normal form (see Appendix \ref{appendix})
\begin{equation}
    y^2 = x(x-1)(x-r),
\end{equation}
where 
\begin{equation}
r = \frac{(a+b)^2}{(a-b)^2 }
\end{equation}
 is the cross-ratio. This explicitly exhibits the curve as a smooth cubic with modular parameter $r$.
 The Legendre form can be further transformed into the Weierstrass normal form
\begin{equation}
y^2 = 4x^3 - g_2 x - g_3,
\end{equation}
where 
\begin{equation}\label{eq:crossratio}
g_2(r) = \frac{4}{3}\big(r^2 - r + 1\big), \quad
g_3(r) = \frac{4}{27}(r - 2)(r + 1)(2r - 1).
\end{equation}
The curve is non-singular if $\Delta=g_2^3 - 27g_3^2=16 r^2(r-1)^2\neq 0$ which is equivalent to $r\neq 0,1$.
The $j$-invariant is
\begin{equation}\label{eq:Jinvariant}
j = 1728 \cdot \frac{g_2^3}{g_2^3 - 27g_3^2}= 256\frac{(r^2-r+1)^3}{r^2(r-1)^2}.
\end{equation}

Using (\ref{SpecEdges}), the cross-ratio is
\begin{equation}
{ \color{red}{r=\frac{(a+b)^2}{(a-b)^2}} }= \frac{1 + \sqrt{1 - 4g} -2g }{2g }.
\end{equation}
Substituting this value of $r$ in (\ref{eq:Jinvariant}), we get
\begin{equation}\label{eq:jofg}
j(g)=\frac{256 (3 g-1)^3}{g^2 (4 g-1)}, \qquad 0<g< \frac{1}{4}.
\end{equation}
We will use this formula to search for values of $g$ that produce a $j$-invariant which is an algebraic integer. We emphasize that the reality of $j$ in (\ref{eq:jofg}) arises because the coefficients $g_2$ and $g_3$ in the Weierstrass form are real functions of the cross‑ratio determined by the spectral edges (see (\ref{eq:crossratio})), which in the Hermitian matrix model lie on the real axis. On the other hand, in non‑Hermitian matrix models the eigenvalues are supported on Jordan arcs in the complex plane, which may lead to a non‑real $j$-invariant.

The reality condition on the $j$-invariant  forces the modular parameter $\tau$ to lie on a one-dimensional skeleton of the fundamental domain $\mathcal{F}_0$. Since the $q$-expansion of $j(\tau)$ has real coefficients, it satisfies the reflection property $j(-\bar{\tau}) = \overline{j(\tau)}$. Consequently, the condition $j(\tau) \in \mathbb{R}$ implies $j(\tau) = j(-\bar{\tau})$. This means that $\tau$ and its reflection across the imaginary axis, $-\bar{\tau}$, must be equivalent under the modular group, i.e., $-\bar{\tau} = \gamma \tau$ for some $\gamma \in \mathrm{SL}(2, \mathbb{Z})$. The only modular transformations capable of mapping a point in $\mathcal{F}$ to this specific reflection are the identity, the inversion $S \colon \tau \mapsto -1/\tau$, and the translation $T \colon \tau \mapsto \tau \pm 1$. These algebraic conditions geometrically constrain $\tau$ to satisfy $\mathrm{Re}(\tau) = 0$, $|\tau| = 1$, or $\mathrm{Re}(\tau) = \pm 1/2$, which correspond precisely to the central imaginary axis, the unit circular arc, and the vertical boundaries of $\mathcal{F}$, respectively, thereby collapsing the two-dimensional domain into a one-dimensional subspace.

Recalling (\ref{eq:algebint}), we focus on the special case $h(D)=1$. In this case, $j$ is not only an algebraic integer, but an  integer, and the complete list of such $j$-values together with their corresponding CM discriminants $D$ is known \cite{lmfdb}. There are thirteen such values in total. However, from (\ref{eq:jofg}), we observe that $j>0$, and thus among these thirteen integer values, our model yields only five admissible cases with $j>0$
\begin{equation}
j_{\text{CM}} = 1728, 8000, 54000, 287496, 16581375.
\end{equation}

Going back to (\ref{eq:jofg}), we notice that 
%\begin{equation}
%\lim_{g\rightarrow 0^+} j(g)\rightarrow \infty,\qquad \lim_{g\rightarrow 1/4^-} j(g)\rightarrow \infty.
%\end{equation}
 $ j(g)$ has a minimum  at $g = 2/9$ with
\begin{equation}
    j(2/9) = 256 \frac{(1 - 3(2/9))^3}{(2/9)^2(1 - 4(2/9))} = 256 \frac{(1/3)^3}{(4/81)(1/9)} = 1728.
\end{equation}
The reason is that $j(\tau)$ has two critical points in the fundamental domain $\mathcal{F}_0$ at $\tau=i, e^{2\pi i/3}$. In our model, only $\tau=i$ is accessible in the domain $g\in (0,1/4)$ and corresponds to $\tau(2/9)=(1+i)/2$ which is $SL(2,\mathbb{Z})$ equivalent to $\tau=i$.

%The value $j = 1728$ corresponds precisely to the square lattice $\mathbb{Z}[i]$, which is the unique elliptic curve (up to isomorphism) with an automorphism group of order 4. 
Furthermore, because the derivative of $j(g)$  vanishes at $g = 2/9$, the point ${\color{red}{g=2/9}}$ is a double root of the equation $j(g) = 1728$. Since $j(g) \to \infty$ as $g \to 0^+$ and $j(g) \to \infty$ as $g \to 1/4^-$, the function $j(g)$ exhibits a strict ``U-shape'' on the physical  $g \in (0, 1/4)$, with a global minimum at $g = 2/9 $.
Because $j(g)$ is a rational function of degree 3, setting $j(g) = j_{\text{CM}}$ for any constant $j_{\text{CM}} > 1728$ yields a cubic equation in $g$. Due to the U-shape of the function on the physical interval, this equation will always yield exactly two distinct real roots, $g_-$ and $g_+$ {\color{red}{within the physical two-cut interval interval $g\in (0,1/4)$}}, such that $0 < g_- < 2/9 < g_+ < 1/4$. 

 We present our results in Table~\ref{tab}. For each admissible $j$-value obtained from our model, we have also listed a corresponding Weierstrass form of the elliptic curve to which the spectral curve is isomorphic over $\mathbb{C}$.
As clearly demonstrated in Table~\ref{tab}, while the smaller root $g_-$ drifts toward $0$, the larger root $g_+$ converges rapidly to $1/4$. For the smallest allowed CM discriminant $D=-28$, the coupling constant is $g_+ \approx 0.249999$, which is indistinguishable from the critical point at standard numerical precision.  The critical coupling $g_c = 1/4$ controls the double-scaling limit, where the two-cut solution collapses into a single-cut phase where the spectral geometry is approaching a singular degeneration when the model begins to exhibit universal critical behavior governed by the Painlev\'e II equation. It is natural that the CM point does not coincide exactly with \(g_c=1/4\). Indeed, at the critical coupling  the  spectral curve degenerates as the two  cuts merge, reducing the curve to genus zero. Since Complex Multiplication is defined only for smooth elliptic curves, the notion ceases to be meaningful precisely at the critical point.

\begin{table}[h!]
\centering
\resizebox{\textwidth}{!}{%
\begin{tabular}{|c|c|c|c|c|}
\hline
CM discriminant $D$ & $j$-invariant &
\multicolumn{2}{c|}{$g=g_{\text{CM}}$} &
Weierstrass Form \\
\cline{3-4}
& & $g_-$ &  $g_+$ & \\
\hline
$-3$   & $0$ & $\times$ & $\times$ &  \\
\hline
$-4$   & $1728$ &  $2/9\approx 0.222222$ & $2/9\approx 0.222222$ & $y^2 = x^3 - x$ \\
\hline
$-7$   & $-3375$ & $\times$ & $\times$ &  \\
\hline
$-8$   & $8000$ &  $0.125000$ & $ 0.247799$  & $y^2 = x^3 +x^2- 13x-21$ \\
\hline
$-11$  & $-32768$ & $\times$ & $\times$ &  \\
\hline
$-12$  & $54000$ & $0.058840$ & $0.249700$ & $y^2=x^3-135x-594$  \\
\hline
$-16$  & $287496$ & $ 0.027778$ & $ 0.249944$ & $y^2=x^3-11x-14$ \\
\hline
$-19$  & $-884736$ & $\times$ & $\times$ &  \\
\hline
$-27$  & $-12288000$ & $\times$ & $\times$ &  \\
\hline
$-28$  & $16581375$ & $0.003891$ & $ 0.249999$ & $y^2=x^3-29155x-1915998$ \\
\hline
$-43$  & $-884736000$ & $\times$ & $\times$ &  \\
\hline
$-67$  & $-147197952000$ & $\times$ & $\times$ &  \\
\hline
$-163$ & $-262537412640768000$ & $\times$ & $\times$ &  \\
\hline
\end{tabular}
}
\caption{Admissible CM discriminants $D$, their corresponding integer $j$-invariants, values of $g_{\text{CM}}$, and the associated elliptic curves in Weierstrass form.}\label{tab}
\end{table}

Consider the case $j = 1728$, corresponding to $\tau = i$ and the square lattice 
$\Lambda = \mathbb{Z} + i \mathbb{Z}$. For $\tau = i$, the lattice is invariant under 
complex multiplication $\lambda \Lambda = \Lambda$ with $\lambda \in \mathbb{Z}[i]$, 
the ring of Gaussian integers, i.e.\ numbers of the form $\lambda = m + ni$ with 
$m,n \in \mathbb{Z}$. This additional symmetry descends to the Weierstrass model 
$E: y^2 = x^3 - x$. Indeed, under the parametrization $(x,y) = (\wp(z), \wp'(z))$ for 
the square lattice, the map $z \mapsto iz$ induces
\begin{equation}
    \wp(iz) = -\wp(z), \qquad \wp'(iz) = i\,\wp'(z),
\end{equation}
where we used
\begin{equation}
    \wp(\lambda z, \lambda \Lambda) = \lambda^{-2}\wp(z,\Lambda), \quad \lambda \Lambda = \Lambda
\end{equation}
Hence, in affine coordinates, 
\begin{equation}
(x,y) \;\mapsto\; (-x,\,iy).
\end{equation}
Together with the involution $(x,y) \mapsto (x,-y)$ arising from $z \mapsto -z$, 
this generates the full automorphism group of $E$, which has order four.

For generic values of the coupling $g$, the spectral curve admits only the involution 
$(x,y)\mapsto(x,-y)$. At the special value $g \approx 0.198$, however, the symmetry is 
enhanced to include $(x,y)\mapsto(-x,\,iy)$, so that the automorphism group has order 
four instead of two.  Analogous symmetry enhancements occur at other CM points 
corresponding to the remaining admissible values of $g$ listed in 
Table~\ref{tab}.

\section{Conclusion}\label{sec:5}

In this work, we have shown how elliptic curves with complex multiplication (CM) naturally arise in the spectral geometry of Hermitian one‑matrix models in the two‑cut phase. For a symmetric quartic potential we derived the genus‑one spectral curve, expressed it in Weierstrass form, and computed its modular $j$‑invariant as a function of the quartic coupling $g$. We identified the special values of $g$ for which the spectral curve acquires CM, leading to enhanced automorphisms at these points.

This provides a concrete link between arithmetic properties of elliptic curves and the spectral data of random matrix ensembles. The analysis extends naturally to higher‑genus spectral curves with $s>2$, where CM becomes a property of the Jacobian variety of the curve, as well as to multi‑matrix models and other ensembles. 
It would also be interesting to analyze the physical implications of CM points, such as their roles in topological recursion or integrable structures.

\section{Acknowledgments}

I am grateful to Ahmed El Gendy and Assem AbdelRaouf for valuable discussions on elliptic curves, and I thank Angese Bissi for  reading the manuscript. I am  grateful to the anonymous referees for their careful reading of the manuscript and for their valuable comments and suggestions.
I would like to acknowledge support through the ICTP-Arab Fund Associates Programme (ARF01 - AFESD Grant No. 14/2023) in the context of the project ``Advancing the Capabilities of Arab Researchers and Students''.

\appendix 
\section{Legendre and  Weierstrass forms}\label{appendix}

In this appendix, we present a detailed derivation of the transformations relating the quartic form of an elliptic curve to its Legendre and Weierstrass forms. Consider an elliptic curve defined by the quartic equation
\begin{equation}
    y^2 = (x-a_1)(x-a_2)(x-a_3)(x-a_4),
\end{equation}
where the branch points $a_1, a_2, a_3, a_4$ are distinct. To transform this model into Legendre normal form, we apply a M\"obius transformation that maps three of the branch points to $0$, $1$, and $\infty$. Define the new affine coordinate \cite{koblitz1993elliptic,silverman2009arithmetic}
\begin{equation}
    X = \frac{(x-a_1)(a_2-a_4)}{(x-a_4)(a_2-a_1)},
\end{equation}
which gives
\begin{equation}
X(a_1) = 0,\qquad X(a_2) =1,\qquad X(a_4)  \mapsto \infty.
\end{equation}
The inverse transformation is 
\begin{equation}
x = \frac{(a_2-a_1)a_4 X - (a_2-a_4)a_1}{(a_2-a_1)X - (a_2-a_4)},
\end{equation}
 with common denominator
\begin{equation} 
 D(X) = (a_2-a_1)X - (a_2-a_4).
 \end{equation}
  Substituting this into each linear factor yields $x-a_i = L_i(X)/D(X)$, where $L_i(X)$ is linear in $X$ for $i=1,2,3$. For $i=4$ the dependence on $X$ cancels exactly, giving
\begin{equation}  
   x-a_4 =\frac{ (a_2-a_4)(a_4-a_1)}{D(X)}, 
\end{equation}     
   so its numerator is constant. Consequently, the product of the four factors reduces to a cubic polynomial:
\begin{equation}
    \prod_{i=1}^4 (x-a_i) = \frac{K \, X(X-1)(X-r)}{D(X)^4},
\end{equation}
where 
\begin{equation}
r = \frac{(a_1-a_3)(a_2-a_4)}{(a_1-a_4)(a_2-a_3)}
\end{equation}
 is the cross-ratio of the ordered quadruple $(a_1,a_2,a_3,a_4)$, and $K \neq 0$ is a constant determined by the $a_i$. The original equation thus becomes
\begin{equation} 
y^2 =\frac{ K X(X-1)(X-r)}{D(X)^4}.
\end{equation}
 Defining a rescaled coordinate $Y = y D(X)^2 / \sqrt{K}$ absorbs the denominator and constant factor, yielding the Legendre normal form
\begin{equation}
    Y^2 = X(X-1)(X-r).
\end{equation}
This explicitly exhibits the curve as a smooth cubic with modular parameter $r$.

To convert the Legendre form to Weierstrass normal form,
we start with the Legendre form of an elliptic curve:
\begin{equation}\label{eq:leg2}
    y^2 = x(x-1)(x-r) = x^3 - (1+r)x^2 + rx.
\end{equation}
Our goal is to transform this into the standard Weierstrass normal form:
\begin{equation}
    Y^2 = 4X^3 - g_2 X - g_3.
\end{equation}

First we {{\color{red}{shift}}} $x$ to remove the quadratic term in the Legendre form (\ref{eq:leg2}). Let $x = X' + c$, substituting this into the expanded Legendre equation yields:
\begin{align}
    y^2 &= (X' + c)^3 - (1+r)(X' + c)^2 + r(X' + c) \\
        &= X'^3 + 3cX'^2 + 3c^2X' + c^3 - (1+r)(X'^2 + 2cX' + c^2) + rX' + rc.
\end{align}
We set
\begin{equation}
    3c - (1+r) = 0 \implies c = \frac{1+r}{3}.
\end{equation}
Substituting $c = \frac{1+r}{3}$ back into the equation, we obtain the depressed cubic:
\begin{equation}\label{eq:depcub}
    y^2 = X'^3 + A X' + B,
\end{equation}
where the new coefficients $A$ and $B$ are:
\begin{align}
    A &= r - \frac{(1+r)^2}{3} = -\frac{r^2 - r + 1}{3}, \\
    B &= \frac{r(1+r)}{3} - \frac{2(1+r)^3}{27} = -\frac{2r^3 - 3r^2 - 3r + 2}{27}.
\end{align}
Multiplying (\ref{eq:depcub}) by  4 gives:
\begin{equation}
    4y^2 = 4X'^3 + 4A X' + 4B.
\end{equation}
We now define our final Weierstrass coordinates $(X, Y)$ as:
\begin{equation}
    X = X' \quad \text{and} \quad Y = 2y.
\end{equation}
Substituting these into the scaled equation yields:
\begin{equation}
    Y^2 = 4X^3 + 4A X + 4B.
\end{equation}
Comparing our result with the standard Weierstrass form $Y^2 = 4X^3 - g_2 X - g_3$, we can directly read off the modular invariants:
\begin{align}
    g_2 &= -4A = \frac{4}{3}(r^2 - r + 1), \\
    g_3 &= -4B = g_3(r) = \frac{4}{27}(r - 2)(r + 1)(2r - 1).
\end{align}
The curve is non-singular if $\Delta=g_2^3 - 27g_3^2=16 r^2(r-1)^2\neq 0$ which is equivalent to $r\neq 0,1$.
The $j$-invariant is
\begin{equation}
j = 1728 \cdot \frac{g_2^3}{g_2^3 - 27g_3^2}= 256\frac{(r^2-r+1)^3}{r^2(r-1)^2}.
\end{equation}

For the symmetric  curve $y^2 = (x^2-a^2)(x^2-b^2)$ used in this paper, the transformation to Weierstrass form is:
\begin{align}
    X &= \frac{(5b^2 - a^2)x + b(5a^2 - b^2)}{3(a + b)^2 (x + b)}, \\
    Y &= \frac{4b(a - b)y}{(a + b)^2 (x + b)^2}
\end{align}
which gives $g_2$ and $g_3$ directly in terms of the spectral edges
\begin{equation}
g_2 = \frac{4(a^4 + 14a^2b^2 + b^4)}{3(a + b)^4}, \qquad g_3 = \frac{8(a^2 + b^2)(a^4 - 34a^2b^2 + b^4)}{27(a + b)^6}.
\end{equation}
Using the values of $a^2 =1/g+2/\sqrt{g}$ and $b^2 =1/g-2/\sqrt{g}$, one can confirm that $g_3(g=2/9)=0$ corresponding to $j=1728$.

\bibliographystyle{plain}

\bibliography{refs}

\end{document}